\documentclass[runningheads,fleqn]{svmult}

\usepackage{makeidx}   
\usepackage{graphicx}  
\usepackage{subeqnar}  
\usepackage{multicol}  
\usepackage{physproc}  
\makeindex             

\begin{document}
\title*{Analysis of the Density of Partition Function Zeroes -
A Measure for Phase Transition Strength}
\toctitle{Analysis of the Density of Partition Function Zeroes -
\protect\newline A Measure for Phase Transition Strength}
%
%
\titlerunning{Analysis of the Density of Partition Function Zeroes}
%
\author{W. Janke\inst{1}
\and R. Kenna\inst{2}
}
%
%
%
\institute{Institut f\"ur Theoretische Physik, Universit\"at Leipzig,
Augustusplatz 10/11, 04109 Leipzig, Germany
\and School of Mathematics, Trinity College Dublin, Ireland
}

\maketitle              

\begin{abstract}
 We discuss a numerical analysis employing the density of partition function
zeroes which permits effective distinction between phase transitions of first
and second order, elucidates crossover between such phase transitions and
gives a new way to measure their strengths in the form
of latent heat and critical exponents. Application to a number of models
demonstrates the efficacy of the technique.
\end{abstract}

%
A central theme of statistical physics is how best to
distinguish between phase transitions of first and second order
from simulational data for finite systems \cite{FSS_reviews}. 
Numerical methods usually exploit the finite-size scaling (FSS) behaviour 
of thermodynamic quantities exhibiting rounded and shifted peaks whose shape 
depends on the order and the strength of the transition and which become 
singular in the thermodynamic limit at the transition point.
An alternative strategy is the analysis of the FSS behaviour of partition 
function zeroes \cite{LY,Fi64,IPZ}. For field-driven phase transitions 
one is interested in the {\em Lee--Yang\/} zeroes in the plane of complex 
external magnetic field $h$ \cite{LY}, and for temperature-driven
transitions the {\em Fisher\/} zeroes in the complex temperature plane are 
relevant \cite{Fi64}. For $d$-dimensional systems below the upper critical 
dimension, the FSS behaviour of the $j^{\rm{th}}$ partition function zero  
(for large $j$) is given by \cite{IPZ}
\begin{equation}
 h_j(L)
 \sim
 \left( j/L^d \right)^{(d+2-\eta)/2d} 
\quad ,
\quad \quad
 t_j(L)
 \sim
 \left( j/L^d \right)^{1/\nu d}
\quad .
\label{compact}
\end{equation}
Here, $L$ is the system size,
$\eta$ is the anomalous dimension, $t = T/T_c - 1$
is the reduced temperature which is zero in the first formula,
and $h$ denotes the external field which is zero in the second formula.
The integer index $j$ increases with distance from the critical point.

%
Despite early efforts \cite{SuKa70}, it has been considered prohibitively
difficult if not impossible to extract the density of zeroes from
finite lattice data \cite{Martin}.
This problem has resurfaced recently as zeroes-related techniques
have become more widespread \cite{KeLa94}.
This provides the motivation for the present work 
in which we
wish to suggest an approach suitable for density analyses.

For finite $L$, the partition function may always
be factorized as
$Z_L(z) = A(z) \prod_{j}{\left(z-z_j(L)\right)}$,
where $z$ stands generically for
an appropriate function of field or 
temperature and $A(z)$ is a smooth non-vanishing function.
Following Suzuki \cite{SuzukiLY} and Abe \cite{Abe}, we assume the 
zeroes, $z_j(L)$,
(or at least those close
to the real axis and hence determining critical behaviour) are
on a singular line for large enough $L$,
impacting on to the real axis at an angle $\varphi$
at the critical point $z=z_c$. The singular line is parameterised by
$z = z_c + r \exp{(i \varphi)}$.
If we define the density of zeroes as
(with $z_j=z_c+r_j \exp{(i \varphi)}$)
\begin{equation}
 g_L(r) = L^{-d} \sum_{j} \delta(r - r_j(L))
\quad ,
\label{J}
\end{equation}
then the cumulative distribution is a a step function,
\begin{equation}
 G_L(r) = \int_0^r{ g_L(s) d s} 
        = j/L^d \quad \quad \quad {\rm{if}} \quad r \in (r_j,r_{j+1})
\quad .
\label{okG}
\end{equation}
It is natural to assume
that at a zero, this distribution function
is given by the average  \cite{LY,GlPr86},
$G_L(r_j) = (2j-1)/2L^d$.

In the thermodynamic limit and for a phase transition of
first order Lee and Yang \cite{LY} showed that the density of zeroes has to 
be non--zero crossing the real axis,
$g_\infty(r) = g_\infty(0) + a |r|^w + \dots$. For the
cumulative distribution this implies the functional form
\begin{equation}
  G(r) = g_\infty(0) r + b|r|^{w+1} + \dots \quad ,
  \label{1st}
\end{equation}
with the slope at the origin being related to the latent heat
(or magnetization) via \cite{LY} $g_\infty(0) \propto \Delta e$.
At second-order phase transitions,
Abe \cite{Abe} and Suzuki \cite{SuzukiLY} have shown that the necessary and 
sufficient condition for the specific heat to have the leading critical 
behaviour $C \sim t^{-\alpha}$, is
$g_\infty (r) = Ar^{1-\alpha}$ or
\begin{equation}
  G(r) \propto  r^{2-\alpha} \quad .
  \label{2nd}
\end{equation}
For $\alpha = 0$, as is the case in the $d=2$ Ising model, it has been 
demonstrated \cite{Abe} that (\ref{2nd}) leads to 
the correct {\em logarithmic\/} divergence in the specific heat.


\begin{figure}[t]
\includegraphics[width=0.45\textwidth,angle=-90,bb=77 65 510 430]{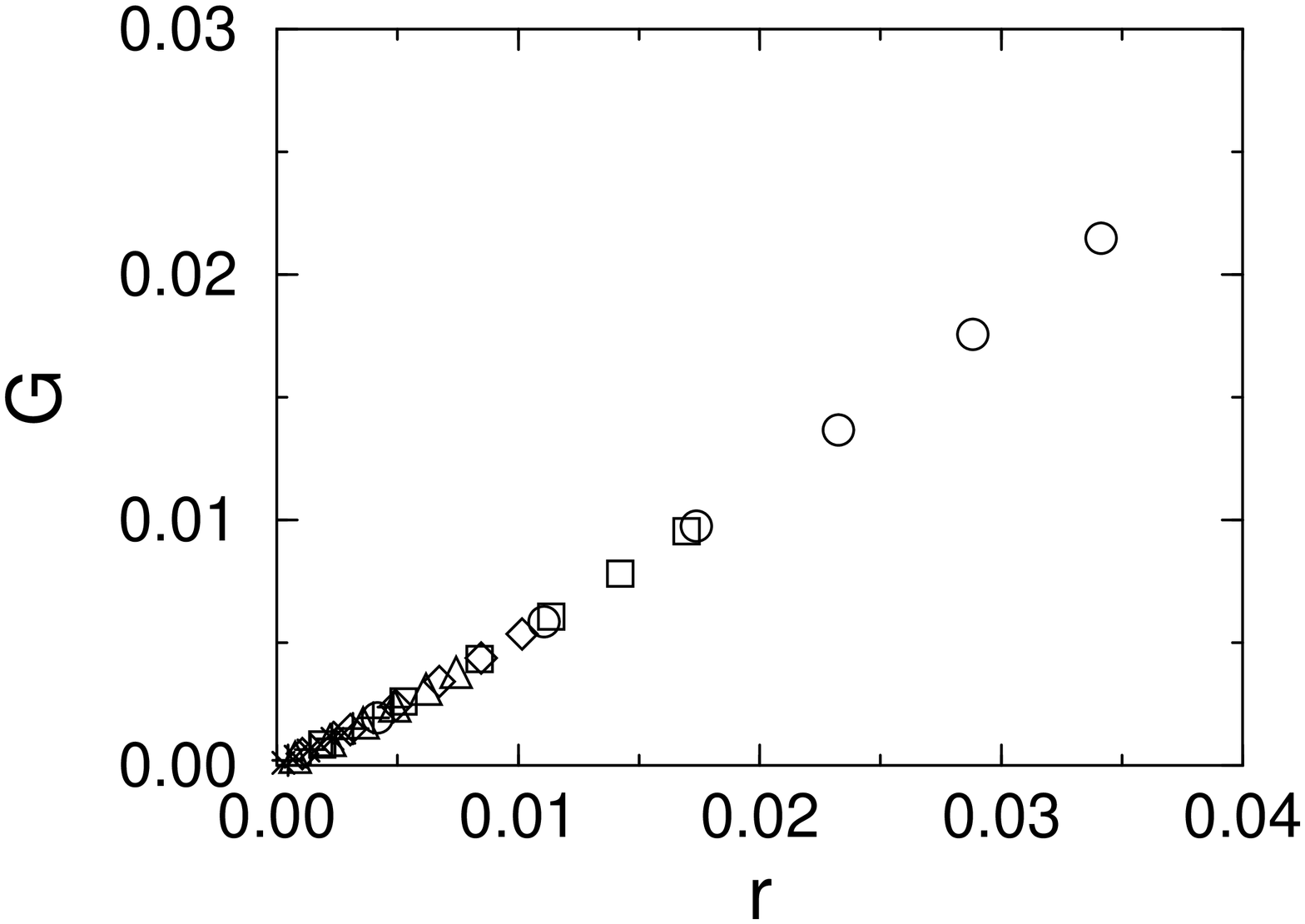}
\vspace*{0.3cm}
\caption[a]{Distribution of the $L=16$--$64$ Fisher zeroes for the 2D 10-state 
Potts model. 
The symbols $\times$, $+,\bigtriangleup,\diamond$,$\put(2.5,0.5){\framebox(4,4)}$~~\,\,, and
$\put(3,2.5){\circle{5}}~\,\,$
correspond to $j=1-6$, respectively.
}
\label{fig:d=2.q=10}
\end{figure}
The preceding considerations show that 
%
%
a plot of $G_L(r_j) = (2j-1)/2L^d$ against $r_j(L)$
should ({\em{i\/}}) go through the origin, ({\em{ii\/}}) display $L$-- and
$j$-- collapse and ({\em{iii\/}}) reveal the order and strength of the phase
transition by its slope near the origin, parameterised generically as
\begin{equation}
 G(r) = a_1 r^{a_2} + a_3 \quad .
\label{gen}
\end{equation}
In order to test the efficiency of the density method we have examined six 
different models \cite{JaKe01} for which various sets of partition-function 
zeroes
exist in the literature: 2D 10-state Potts, 3D 3-state Potts, 3D and 2D Ising, 
(3+1)D SU(3), 4D Abelian Surface Gauge. Here we shall summarize the results 
for the first three of these models.

\paragraph{2D 10-state Potts model:} 
The two-dimensional $q$-state Potts model is the classic testing ground for 
analytical and numerical studies of first-order ($q > 4$) and second-order 
($q\le 4$) phase transitions. Apart from the transition point,
$\beta_0 = \ln(1+\sqrt{q})$, also the critical exponents ($q\le 4$), the
latent heat ($q > 4$) and various other moments at $\beta_0$ are 
known exactly.

Our analysis of the density of zeroes as tabulated in Refs.~\cite{Vi91,ViAl91} 
begins with Fig.~\ref{fig:d=2.q=10} where the first six Fisher zeroes are 
plotted for $L=16$--$64$. 
We observe 
excellent $L$-- and $j$-- collapse indicating that $G_L(r_j) = (2j-1)/2L^d$
is the correct functional form.
Fitting (\ref{gen}) to the $j=1$--$4$ points gives
$a_2=1.10(1)$, $a_3= 0.00004(1)$,
strongly indicative of a first-order phase transition.
Indeed, fixing $a_3=0$, and fitting the two remaining parameters
to the lowest four
data points gives $a_2=1.008(6)$. Assuming thus $a_2=1$, $a_3=0$, 
and applying a single-parameter fit to the full data set
we obtain $g(0) = a_1 = 0.501(8)$.
Further fits close to the origin
yield the slopes and corresponding estimates for the latent heat
$\Delta e = 2 \pi g(0) \exp(-\beta_0)$ 
indicated in Table~\ref{tab:210}.

\begin{table}[b]
\caption{Fits of the cumulative distribution $G(r)$ 
to the $N$ lowest Fisher zeroes ($L=16$--$64$, $j=1$--$4$) of the 2D 10-state 
Potts model.}
\label{tab:210}
\begin{tabular*}{\textwidth}{@{\extracolsep{\fill}}llllllll}
\hline
\multicolumn{1}{c}{$N$}  &
\multicolumn{1}{c}{24}   &
\multicolumn{1}{c}{16}   &
\multicolumn{1}{c}{12}   &
\multicolumn{1}{c}{8}    &
\multicolumn{1}{c}{4}    &
\multicolumn{1}{c}{exact}    \\
\hline
$g(0)$    & 0.501(8) 
& 0.479(3) & 0.471(2) & 0.469(2) & 0.463(1) & 0.4611\\
$\Delta e$& 0.756(11)
& 0.723(4) & 0.711(3) & 0.708(2) & 0.698(2) & 0.6961\\
\hline
\end{tabular*}
\end{table}

\paragraph{3D 3-state Potts model:}
\begin{figure}[t]
\includegraphics[width=0.45\textwidth,angle=-90,bb=77 65 510 430]{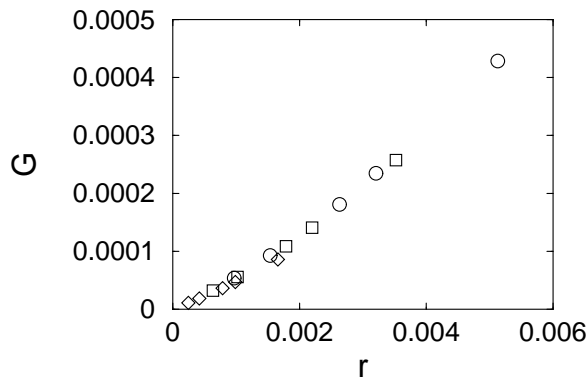}
\caption[a]{Distribution of the $L=18$--$36$ Fisher zeroes for the
3D 3-state Potts model. 
The symbols $\diamond$,$\put(2.5,0.5){\framebox(4,4)}$~~\,\,, and  
$\put(3,2.5){\circle{5}}~\,\,$
correspond to $j=1,2$, and $3$, respectively.
}
\label{fig:d=3.q=03}
\end{figure}
It is generally accepted that this model exhibits a first-order phase 
transition, albeit a very weak one \cite{JaVi97}. This is therefore
a typical model used to test new methods to discriminate between
first- and second-order transitions.
A list of the first few Fisher zeroes (with $z=\exp{(-3\beta /2)}$) 
for $L=10-36$ can be found in Refs.~\cite{Vi91,ViAl91}.

Our density analysis is presented in Fig.~\ref{fig:d=3.q=03}.
A 3-parameter fit to all data yields $a_3=0.000005(2)$
and becomes even closer to zero as the fit is restricted closer
to the origin.
Clearly the slope is non-zero near the origin -- the signal of a 
first-order phase transition. In fact, a 2-parameter fit to
the data corresponding to $L=22,24,30,36$, $j=1$ yields
$a_2=  1.06(2)$. Accepting that the plot
is in fact linear near the origin, and fitting for the slope only gives
$g(0) = a_1 = 0.0454(9)$.
Using $\beta_0=0.3670$ \cite{Vi91,ViAl91,JaVi97}
and $\Delta e = 2 \pi g(0) (3/2) \exp(-3\beta_0/2)$,
we find that the corresponding latent heat is $\Delta e = 0.247(5)$,
comparing well with $0.2409(8)$ from \cite{Vi91,ViAl91} and with $0.2421(5)$ 
from the more sophisticated analysis of \cite{JaVi97}.
\vspace*{-0.3cm}
\paragraph{3D Ising model:}
The first seven exact Fisher zeroes in the $z=\exp({-4\beta})$ plane
for $L=4$ are given in \cite{Pe82},
together with numerically determined zeroes for $L=5$, $j=1-4$.
We also use the zeroes in Refs.~\cite{Vi91,Ma84} for $L=7$, $j=1,2$; 
$L=6,8,10,14$, $j=1-3$; $L=32$, $j=1$. 

\begin{figure}[t]
\includegraphics[width=0.42\textwidth,angle=-90,bb=77 65 510 430]{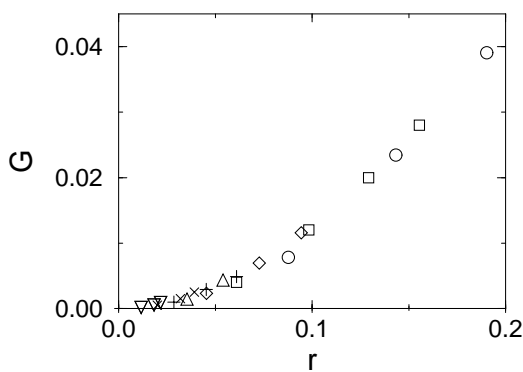}
\vspace*{0.1cm}
\caption[a]{Distribution of the $L=4$--$32$ Fisher zeroes for the
3D Ising model. 
The symbols $\times,+$,$\put(2.5,0.5){\framebox(4,4)}$~~\,\,, 
and $\put(3,2.5){\circle{5}}~\,\,$
correspond to $j=1,2,3$, and $4$, respectively.
}
\label{fig:d=3.q=02}
\vspace*{-0.1cm}
\end{figure}

Fitting the ansatz (\ref{gen}) to the full set of $L=4-32$, $j=1-3$ data
shown in Fig.~\ref{fig:d=3.q=02}
indicates a second-order phase transition with $a_2=1.81(3)$, 
$a_3 = -0.00001(1)$. Accepting $a_3=0$ and applying a 2-parameter
fit to the six data points corresponding to $L=10-32$, $j=1$,
gives $a_2= 1.879(2)$ or $\alpha = 0.121(2)$, roughly 
compatible with the weighted ``world average'' \cite{JaWe00} of
$\alpha = 0.10985(54)$.

\vspace*{0.3cm}
%
%
To summarize, we have shown that from the qualitative behaviour of the
cumulative density of partition function zeroes we can distinguish between 
first- and second-order transitions while from the quantitative details we 
can extract the latent heat and the specific-heat exponent $\alpha$,
respectively. Our method meets with a high degree 
of success even in the borderline case of the 3D 3-state Potts model
where with traditional methods the distinction between a first- and 
second-order phase transition is quite difficult. 
%


%

\clearpage
\addcontentsline{toc}{section}{Index}
\flushbottom
\printindex

\end{document}